%% file: main.tex
\documentclass[sigconf]{acmart}
\renewcommand\footnotetextcopyrightpermission[1]{}
\usepackage{tikz}
\usepackage{amsmath}
\usepackage{xspace}
\usepackage{multirow}
\usepackage{xcolor,colortbl}
\usepackage{tabularx}
\usepackage{booktabs}
\usepackage{xcolor}
\usepackage{listings}
\usepackage{subcaption}
\usepackage{xparse}

\usepackage{makecell}

\usepackage[bb=boondox]{mathalfa}
\usepackage{amsmath}
\usepackage{bm}
\usepackage{tikz}
\usepackage{pgfplots}
\pgfplotsset{width=7.5cm,compat=1.12}
\usepgfplotslibrary{fillbetween}
\pgfplotsset{
    /pgfplots/layers/Bowpark/.define layer set={
        axis background,axis grid,main,axis ticks,axis lines,axis tick labels,
        axis descriptions,axis foreground
    }{/pgfplots/layers/standard},
}

\usepackage[normalem]{ulem}
\usepackage{algorithm}
\usepackage{algorithmicx}
\usepackage{algpseudocode}
\usepackage{amsmath}
\usepackage{tcolorbox}

\algnewcommand{\LineComment}[1]{\State \(\triangleright\) #1}
\algdef{SE}[DOWHILE]{Do}{doWhile}{\algorithmicdo}[1]{\algorithmicwhile\ #1}%
\definecolor{Gray}{gray}{0.9}

\begin{document}
\title{
A Survey of Prevent and Detect Access Control Vulnerabilities
}

\author{Li Zhong}
\affiliation{%
  \institution{University of California, San Diego}
  \country{United States}}
\email{lizhong@eng.ucsd.edu}

\input{abstract}
\maketitle
\pagestyle{plain}
\input{intro}

\input{background}

\input{overview}
\input{Design-Implementation}
\input{Analysis-Testing}
\input{Runtime-Monitoring}
\input{discussion}
\input{conclusion}

\clearpage
\bibliographystyle{ACM-Reference-Format}
\bibliography{ref}

\end{document}

%% file: abstract.tex
\begin{abstract}
Broken access control is one of the most common security vulnerabilities in web applications. These vulnerabilities are the major cause of many data breach incidents, which result in privacy concern and revenue loss. However, preventing and detecting access control vulnerabilities proactively in web applications could be difficult. Currently, these vulnerabilities are actively detected by bug bounty hunters post-deployment, which creates attack windows for malicious access. To solve this problem proactively requires security awareness and expertise from developers, which calls for systematic solutions.

This survey targets to provide a structured overview of approaches that tackle access control vulnerabilities. It firstly discusses the unique feature of access control vulnerabilities, then studies the existing works proposed to tackle access control vulnerabilities in web applications, which span the spectrum of software development from software design and implementation, software analysis and testing, and runtime monitoring. At last we discuss the open problem in this field.
\end{abstract}

%% file: intro.tex
\section{Introduction}
\label{sec:intro}
Nowadays web services have became an important part of our life and preserve huge amount of sensitive and private data of their users. To protect these data, web services need to maintain access control inside the system, which typically implemented as authorization checks inside the application code. However, to get things right is not easy. From decades ago, broken access control has become a severe problem in web services. Missing authorization and improper authorization were ranked 6th and 15th respectively in 2011 CWE/SANS report as most dangerous software vulnerabilities~\cite{2011-stat}. And it even gets worse these days -- in the recent year 2021, the OWASP ranked broken access control as the first among top ten web application risks, with the most occurrences in real-world applications~\cite{2021-stat}. Some recent security news are also highly related to such vulnerabilities, as shown in Table~\ref{tab:news}. 

\begin{table*}[!hbt]
\begin{tabular}{|l|l|l|}
\hline
\textbf{Website}                  & \textbf{Description}                                                                                 & \textbf{Time}                      \\ \hline
\textbf{Jared and Kay Jewelers\cite{jared}}   & Exposed order information including customers' name, address, phone, email, etc & December 2018 \\ \hline
\textbf{LifeLock\cite{lifelock}}                 & Exposed customers' email addresses.                                                                  & July 2018                          \\ \hline
\textbf{Panera Bread\cite{panerabread}}              & Exposed customers' name, email, address, birthday, last four digits of credit card            & April 2018                         \\ \hline
\textbf{Facebook\cite{fb-bug}}                 & Allowed anyone to delete other people's photos                                                       & February 2015                      \\ \hline
\textbf{First American Financial\cite{faf}} & Leaked more than 800 million real estate documents                                                   & June 2021                          \\ \hline
\textbf{JustDial\cite{justdial}}                 & Leaks 100 million users' names, emails, phone numbers, etc                                           & Apr 2019                           \\ \hline
\textbf{AOL\cite{aol}}                      & Leaked all search queries and reactions of its users                                                 & August 2006                        \\ \hline
\end{tabular}
\caption{Security accidents caused by broken access control in real world websites}
\label{tab:news}
\end{table*}

Unfortunately, it is still difficult to prevent and detect such vulnerabilities in the development of web applications. The difficulty comes from multiple aspects. 
On the one hand, modern web applications are widely written in framework that are not natively put access control as first-class role, like Django~\cite{django} and Ruby on Rails~\cite{ruby}. 
As a result, most web applications implement its access control policy in an ad-hoc manner throughout the application code. 
On the other hand, access control vulnerabilities are hard to tested and detected in development. 
Unlike other vulnerabilities, e.g. SQL injection or XSS attacks, which has explicit symptoms such as irregular SQL queries or abnormal data flows, access control vulnerabilities usually flow legal control and data flow inside the system. 
Whether an access is unauthorized is determined by the internal logic of a web application. 
These factors contribute to the surprising fact that this problem has long lasting and still cause big security accidents until now. 

As the consequence of access control vulnerabilities could be very severe, a lot of work are proposed to tackle with this problem.
These work span across the development process of a web application and solve this problem from different aspects.
Some works help to detect such vulnerabilities before the systems are deployed, while some works monitor the runtime execution and detect skeptical accesses to defense against them after the systems are deployed. Some work participate in the first place that help developers to prevent access control vulnerabilities. They either provide centralized access control frameworks~\cite{cancan, cancancan, caslin}, or propose new access control mechanisms based on architecture design, language support and database support.

In this survey, we provides a structured overview on the existing approaches on preventing and detecting access control vulnerabilities in web applications in different phase of development cycles. We mainly focus on the vulnerabilities introduced by developers of web applications that contain different roles, permissions and resources in their business logic. We hope that with this survey, we can have a better understanding on this problem and overview what existing solutions are proposed for this problem. The rest of this survey are organized as follows: In Section~\ref{sec:background}, we begin with a description on the problem scope, definitions and the challenges to understand the problem. Then in Section~\ref{sec:overview}, we provide an overview of the state-of-the-art approaches that tackle with this problem in different phases of software development. In Section~\ref{sec:tool}, \ref{sec:analysis} and \ref{sec:monitor}, we will respectively discuss the details of tool support in software design and implementation, software analysis and testing and runtime monitoring. In Section~\ref{sec:discuss}, we discuss the new opportunities and open problems with regard to access control vulnerabilities.

%% file: background.tex
\section{Preliminary:Access Control and Access Control Vulnerabilities}
\label{sec:background}
\subsection{Access Control}

In web applications, access control is the mechanism than restricts operations on resources. Nowadays software is used to perform access control by the workflow of authentication and authorization. These two concepts are separate processes and cannot be used interchangeably. Therefore, we clarify their definitions here and further talk about the definitions and features of access control vulnerabilities.

\paragraph{\textbf{Authentication}} Authentication is the process of identifying someone is who they claim to be. This process can work through many ways, including passwords, one-time pins, biometric verification, etc. These authentication approaches can be combined to further increase the difficulty of cracking them all, thus increasing the confidentiality. Since most of web applications are using HTTP/HTTPS stateless  protocol, it means they cannot maintain the login state like how the access control mechanism does in UNIX/Linux systems. To make up for this, web applications use security tokens like cookies and session tokens to maintain the authenticated identity of current users so they can avoid annoying process of repeating login.

\paragraph{\textbf{Authorization}} Authorization is the process inside a system to decide what access a user is granted. It always takes place after authentication. Without special features, authorization settings in a web applications are usually implemented and maintained by the organizations using this system by staffs with special roles like administers, or the core developers, which is different from authentication for which even normal users may have options to change the token/secrets. 

Therefore, in a web application, the typical workflow is as follows:
\begin{itemize}
    \item A user logs in and the server returns a generated randomized token, which can be transmit between clients and servers in the following requests as credentials. (Authentication)
    \item The client includes that token in the requests, which identify himself as the login users. Then server decide whether the permission required in this request is satisfied. (Authorization)
    \item The server responds to the user with either an error page due to insufficient permission, or some other information based on the user's permissions. (Authorization)
\end{itemize}

However, authentication is not a mandatory precondition for authorization. For example, an unauthenticated user could be authorized to public resources of a web application even before he identifies himself. There is another exception case where the implementation of authentications and authorizations are combined and simplified-- for example, in collaboration platforms like Overleaf, Google Doc, they contain a long randomized string in the URL and anyone who get the URL can access to the content. The authorization process is shortcut as `anyone get the randomized token embedded in the URL is regarded as having the capability of access something'. The management of such tokens embedded into URLs is up to the users who get them.

\subsection{Access Control Vulnerabilities}
\label{subsec:ac-vulnerabilities}
Access control vulnerabilities enable the attacker to access web pages containing sensitive or private data and perform unauthorized actions. It happens when the intended access control policies are not enforced inside the web applications, which could end up in severe consequence like data leakage, Quality of Service (QoS) degradation, and causes financial loss to the service providers. 

The categories of access control vulnerabilities can be divided into three: vertical, horizontal and context-dependent access control vulnerabilities. 
\begin{itemize}
    \item \textbf{Vertical Access Control Vulnerabilities.} This allow users to illegally access the resources of another user with totally different roles. A typical example is normal users may have access to resources that should have been only accessed by administrators.
    \item \textbf{Horizontal Access Control Vulnerabilities.} It happens when a user can illegally access resources of another user with the same role. For example, in a email applications, a user should only access to mails of himself. If he can access mails of another user, that is a horizontal access control vulnerabilities.
    \item \textbf{Context-dependent Access Control Vulnerabilities.} It happens when some resources should only have been accessed only in multiple stages within a process. For example, when a customer buys an item from a E-Commerce web applications, he should go though the payment step before he obtain an item confirmation page. If that context dependency is broken, it will result in an context-dependent access control vulnerabilities.
\end{itemize}
 
\begin{figure}[!hbt]
    \includegraphics[width=0.5\textwidth]{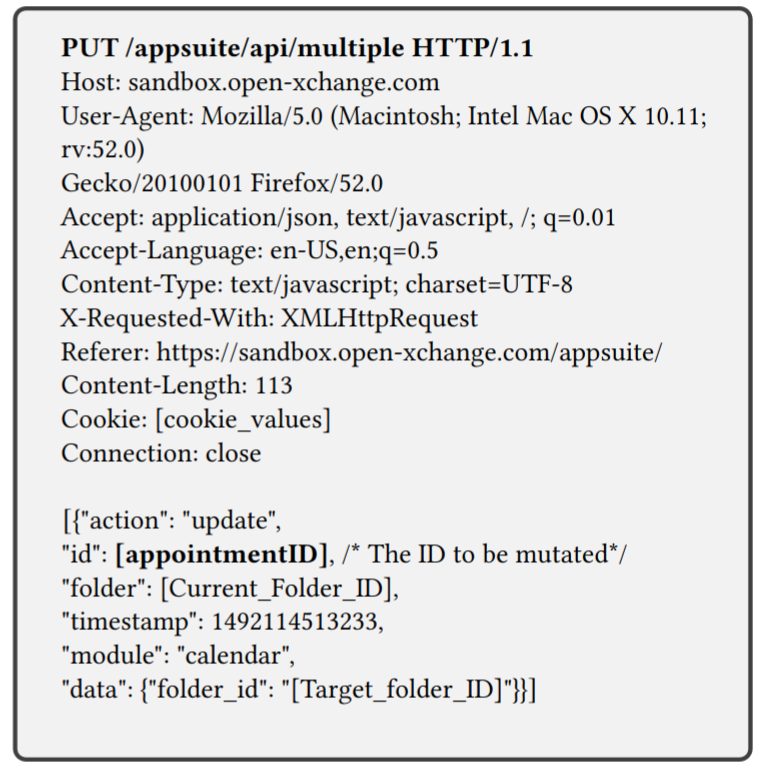}
    \caption{A request that can trigger access control vulnerability in Open-Xchange.}
    \label{fig:open-xchange}
\end{figure}

To exploit such vulnerabilities, there are multiple ways. If the attacker try modify parameters inside his request to make the identifiers point to other sensitive resource, this is referred as parameter manipulation\cite{li2013logicscope}. If he tries to access a restricted page by directly access by the URL of that page, this is referred as forceful browsing~\cite{force-browse}.  All these attacks try to find the holes in access control policy enforcement and bypass the access control. Figure~\ref{fig:open-xchange} show such an example. In the real world case from Open-Xchange~\cite{ac-example}, by mutating the "id" field in the payload of the PUT request, an attacker, as a normal user that has no difference from other users, could access private appointments that belong to other users. The vulnerability wins bug bounty award and gets fixed soon after the report, which reveals the importance of this kind of vulnerabilities.

\subsection{What Causes Access Control Vulnerabilities}

Access control vulnerabilities can be introduced from different ways. By the location of vulnerability root causes, we can divide them into two different type: code logic error and misconfiguration.

\textit{Code Logic Error} Since the access control checks in web applications are usually implemented in an ad-hoc way nowadays, it is hard to distinguish between the concrete code patterns. However, from how these errors are exploited, the code logic error can happen in following cases: 1.  Unprotected endpoints: sensitive functionality and resources can be accessed without appropriate protection, like the administrative endpoints `https://example.com/admin/'. Making the URL that grant access to be less easy to guess could help, but a cryptic URL does not guarantee security even it is not referred inside the application to low-privileged users. 2. Identifier-based endpoints: by mutating the identifiers in requests, attackers could gain access to other resources without protection, as we show in Figure~\ref{fig:open-xchange}. 3. Multistage endpoints: for process that involves multiple stages, any step that fails to validate the request sender's identity could result in potential vulnerabilities. A malicious user could intercepts the request and modify the parameters to perform unauthorized operations.

\textit{Misconfiguration} Web applications could also have access control on application platform layer to control which user perform what HTTP methods (GET, POST, etc.) on specific URL paths. Besides, static files in a web application are also protected by web server configurations, which are usually located within the server's web root. Since no applications-level code intercepts in their handling, the static resources can be accessed directly in this way. The vulnerabilities caused by misconfigurations are discussed in studies on misconfigurations~\cite{xu2015systems, xu2015hey}. In this survey we focus mainly on access control vulnerabilities caused by developers' mistakes in implementation.

\subsection{Why Access Control Vulnerabilities Are Hard to Deal With}

The difficulties of eliminating access control vulnerabilities from web applications are rooted in their semantics, complexity, dynamism and the lack of systematic support for enforcement and testing.

\textit{Semantics} Unlike other vulnerabilities such as SQL injection and cross site scripting (XSS)~\cite{shar2012mining, zhang2019machine, halfond2006using} where the attackers need make up special input into the web applications, access control vulnerabilities usually manifests with no such abnormal input. They are difficult to prevent by data sanitization even though access control vulnerabilities are also related to data flow vulnerabilities. Besides, access control policies of a web application are deeply bound to the business logic of this application. And the legality of access is validated by developers' intention, without a fixed ground truth. The semantics of access control makes this problem difficult to tackle with, which calls for the collaboration from developers.

\textit{Complexity} Nowadays web applications, with the development of IT industry, are no longer easily fitting into simple access control models like DAC, RBAC or MAC. The permissions or capabilities are divided into finer granularity and deeply bound to the application logic. In popular web applications like Facebook, the access control policy can even be continuously updated by users in the privacy settings~\cite{paul2011improving, egelman2011oops}, which means even the capabilities on same type of resources should be separately handled. Implementing access control mechanisms in correct ways is also not an easy task. As ~\cite{wang2013explicating} pointed out, untrained developers could ignore some underlying assumptions on their implementations. \cite{zuo2017authscope} also considers that not all of the online service providers follow the best practice when using security tokens, and they sometimes does not enforce the check on token belongings. In another case, Orange CERT~\cite{orange} use short randomized tokens that can be easily exploited. Due to the unawareness and lack of security trainings, it is hard for developers to enforce access control mechanism in the correct way.

\textit{Dynamism} New objects and subjects are added constantly to the applications. And the interactions between users could also result in the grant or revoke of new permissions. Therefore, in web applications, access control policy in web applications are continuously keeping updated, which leave the attackers with information that could lead to exploitations, like the resource identifiers, URL patterns, etc. Besides, one unique characteristic of web applications is that the functionalities in them are keep changing. Old deprecated functionalities leave outdated APIs~\cite{yasmin2020first} and orphaned pages~\cite{pletinckx2021out} if not getting attention, while new added features could be vulnerable to access control bypass without thorough testing.

\textit{Lack of Systematic Support} The consequence of broken access control could be largely different than other broken logic code, as we discuss in Section~\ref{subsec:ac-vulnerabilities}. Most broken code can be fixed and return to normal use, while broken access control could result in irreversible data leak and affect the victims even after the patches. 
Due to this special attribute, the tool support should be more on the pre-deployment phase as proactive ways.
Current popular web frameworks, like Django, Ruby on Rails, Vue.js, emphasize on fast and smooth development on functionalities. Access control are treated as part of the functionalities that implemented by developers instead of an enforcement. But this problem does not get special treatment in popular web frameworks, which usually follows the software architectural patterns of MVC~\cite{deacon2009model}, MVP~\cite{mvp} or MVVM~\cite{mvvm}. Similarily, the testing support for access control in web applications is insufficient. In most web applications, there is not dedicated access control testing. Instead, the access control tests are embedded in other functionality testing as an intermediate steps. This attitude hinders a thorough testing to cover all the objects, subjects and operations inside the system while access control vulnerabilities hide in those corner cases that few tests can cover.

%% file: overview.tex
\section{Overview}
\label{sec:overview}
\begin{table*}[!hbt]
\begin{tabular}{l|l|l|l|l}
\hline
\textbf{Stage} & \textbf{Category} & \textbf{Human Effort} & \textbf{Coverage} & \textbf{Accuracy} \\ \hline
\multirow{4}{*}{\textbf{\begin{tabular}[c]{@{}l@{}}Tool Support in\\ Software Design \\ and Implementation\end{tabular}}} & \textbf{Framework Support} & High & High & \multicolumn{1}{c}{-} \\ \cline{2-5} 
 & \textbf{Database Support} & High & High & \multicolumn{1}{c}{-} \\ \cline{2-5} 
 & \textbf{Language Support} & High & High & \multicolumn{1}{c}{-} \\ \cline{2-5} 
 & \textbf{\begin{tabular}[c]{@{}l@{}}Development Environment \\ Support\end{tabular}} & Low & High & Low \\ \hline
\textbf{\begin{tabular}[c]{@{}l@{}}Software Analysis \\ and Testing\end{tabular}} & \textbf{Outlier Detection} & Low & High & Low \\ \cline{2-5} 
\textbf{} & \textbf{Automated Testing} & Low & Low & High \\ \cline{2-5} 
\textbf{} & \textbf{Formal Method} & Low - High & High & Low \\ \hline
\textbf{\begin{tabular}[c]{@{}l@{}}Deployment \\ and Monitoring\end{tabular}} & \textbf{\begin{tabular}[c]{@{}l@{}}Runtime Monitoring \\ and Tracking\end{tabular}} & High & Low & High \\ \hline
\end{tabular}
\caption{The Overview of Existing Approaches to Prevent and Detect Access Control Vulnerabilities}
\label{tab:category}
\end{table*}
In this survey, we discuss the state-of-the-art approaches that are proposed to tackle with access control vulnerabilities. Since in this survey, we target at the vulnerabilities in application code, based on the typical software development life cycle, we divide these approaches across the stages including software design and implementation, software analysis and testing, deployment and monitoring. Table~\ref{tab:category} and Table~\ref{tab:overview} gives an overview of these categories and the related works in each stage. We hope to help both developers and researchers in this field to understand the existing solutions in this field, and encourage more people to fill in the blank and push this problem forward.

\subsection{Tool Support in Software Design and Implementation}
Software design and implementation is the stage when an executable software system is developed. With customized requirements, developers need to identify components and relationships in the software and realize the software design in specific languages and frameworks, which end up with the executable software. In this process, the design and implementation choice could affect the possibility of introducing access control vulnerabilities into this software systems. A series of work are done in this line to help eliminate the access control vulnerabilities from the root, which by the components they locate can by divided into support in framework level, database level and language level. Besides, a special component in software design and implementation, the development environment, can also serve as helpers in tackling with access control vulnerabilities.

\paragraph{Framework Support} As the mainstream web framework like Django and Ruby on Rails do not put security concerns as the first-class considerations, some modules and libraries are proposed to supplement the requirements on security goals as an extension to existing frameworks. Besides, many previous works also notice the internal drawbacks of the widely used MVC frameworks, thus propose to enforce access control policies in the applications by providing interface of domain-specific language (DSL) alongside the data models. With the enforcement, it is possible to put all data flow under mandatory access control (MAC), which will eliminate some access control vulnerabilities that totally ignore security checks. We discuss the framework support for prevent access control vulnerabilities in Section~\ref{sec:framework}.

\paragraph{Database Support} As the source and destination of data operations, access control support from database is explored to restrict access control vulnerabilities from being exploited. One direction in DBMS layers is to take advantage of existing database features like SQL conditions and database views, while some other work design special databases and middleware that internally enforce access control. The access control policy could either from developer's specifications, or from learning and inference on past queries. We discuss the database support for prevent access control vulnerabilities in Section~\ref{sec:database}.

\paragraph{Language Support} The best way to prevent access control vulnerabilities is to avoid introducing it in the first place. In software development, one of the fundamental consideration is the programming language developers will use. Therefore, solutions on programming language layer are proposed to enforce access control policies at the early stage. One direction is to enforce information flow control so that if data flow must obey the security assertions/attributes on data objects. This enforcement can be achieved by taking advantage of programming language concepts and techniques like refinement types, logical attestations, static information flow analysis, etc. We discuss the language support for prevent access control vulnerabilities in Section~\ref{sec:language}.

\paragraph{Development Environment Support} While all the supports we discussed above could require additional effort from developers to written access control policies in special languages or APIs, a more approachable way to tackle with this problem without bringing much additional effort is to remind them of potential access control vulnerabilities in development environment. This solution is non-intrusive and interactive, guaranteeing the participation from developers since the access control policy is more of a semantic problem, but with less effort. Not every developers are motivated to switch to non-mainstream databases, languages or frameworks with special access control support, which could potentially lack of community support. But the learning cost of switching to another IDE could definitely help with this problem more practically. We discuss the development environment support for prevent access control vulnerabilities in Section~\ref{sec:development}.
\subsection{Software Analysis and Testing}

Aspects of software development besides programming comprise over 50\% of development costs~\cite{computing2003grand}, among which software analysis and testing serve an important role. Software analysis discover facts about a given program, while software testing conduct with both manually written and automatically generated test cases to evaluate and verify the behavior of the software. Despite the aforementioned efforts, access control vulnerabilities as human errors are still inevitable~\cite{patterson2002recovery}. Software analysis and testing serve as the second defense to this devil. 

\begin{figure}
    \centering
    \includegraphics[width=\linewidth]{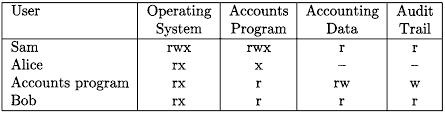}
    \caption{An example of access control matrix.}
    \label{fig:ac-matrix}
\end{figure}

\paragraph{Outlier Detection}
According to Clark-Wilson model~\cite{blake2000clark}, access control model can be expressed as a relationship between an authenticated principal (subject), a set of programs (program), and a set of data items to be operated on (object), namely access control triple. Butler Lampson, in his 1971 paper \textit{Protection}~\cite{lampson1974protection}, proposed access control matrix as well to abstract security model of access control. Figure~\ref{fig:ac-matrix} shows such an example. As an intuition, outliers or deviants in rows and columns of access control models could indicate a signal of violations. Outliers detection relies on the construction of `inliners', which require a base criteria. This criteria comes from two source: developer specification and policy inference. A number of tools are proposed based on this workflow, which we will discuss in Section~\ref{sec:outlier}.

\paragraph{Automated Testing} 
Many previous work of access control policy testing assumes there is a dedicated component to control over the access to resources, which is normally called the policy decision point (PDP). And the policies typically are written in XACML by administrations, making it more of a misconfiguration problem instead of code logic bugs. However, in our survey, we do not place such assumptions on most of the web applications. Therefore, we discuss the related works that target to test access control vulnerabilities in web applications implementation. These works range from white-box to black-box testing, which we will discuss in Section~\ref{sec:testing} in detail.

\paragraph{Formal Method} Formal methods use rigorously mathematical models to ensure correct behaviors of systems, which serve as a complement to system testing. It uses formal verification schemes to prove the correctness of a system. In access control scenario, formal methods are used to validate the correctness of access control implementation correctness in this applications, based on the security models either proposed by developers or learned from execution traces. We discuss these approaches in Section~\ref{sec:formal}.

\subsection{Runtime Monitoring}
Runtime monitoring and tracking serves as the last defense to illegal access. These approaches, compared to testing and analysis, have the advantages of adapting to evolving web application languages and frameworks. While the technique trend keeps changing, most web applications still rely on HTTP protocol for transmitting information. Therefore, monitoring provides a promising one-size-fits-all solution. Also, runtime monitoring approaches allow for adaption to the dynamism of access control policies. The monitor agent can keep updating the intended policies it validates against. We discuss these monitoring approaches in Section~\ref{sec:monitor}.

\begin{table*}[]
\begin{tabular}{l|l|l}
\hline
\textbf{Stage} & \textbf{Category} & \textbf{Representative Works} \\ \hline
\multirow{4}{*}{\textbf{\begin{tabular}[c]{@{}l@{}}Tool Support in\\ Software Design \\ and Implementation\end{tabular}}} & \textbf{Framework Support} & CanCan, Casbin, Diesel, OKWS, Passe, CLAMP, W5, Hails \\ \cline{2-3} 
 & \textbf{Database Support} & David et al., Gabriel et al., Shariq et al., DataLawyer, Qapla, \\ \cline{2-3} 
 & \textbf{Language Support} & Jif, Jeeves, Jacqueline, LWeb, RESIN, Storm, Hails \\ \cline{2-3} 
 & \textbf{\begin{tabular}[c]{@{}l@{}}Development Environment \\ Support\end{tabular}} & ASIDE \\ \hline
\textbf{\begin{tabular}[c]{@{}l@{}}Software Analysis \\ and Testing\end{tabular}} & \textbf{Outlier Detection} & MACE, SPACE, RoleCast, Sun et al., Waler, CanCheck, FixMeUp \\ \cline{2-3} 
\textbf{} & \textbf{Automated Testing} & AuthScope \\ \cline{2-3} 
\textbf{} & \textbf{Formal Method} & LogicScope, Wang et al., \\ \hline
\textbf{\begin{tabular}[c]{@{}l@{}}Deployment \\ and Monitoring\end{tabular}} & \textbf{\begin{tabular}[c]{@{}l@{}}Runtime Monitoring \\ and Tracking\end{tabular}} & Nemesis, FlowWatcher \\ \hline
\end{tabular}
\caption{The Overview of Existing Approaches to Prevent and Detect Access Control Vulnerabilities}
\label{tab:overview}
\end{table*}

%% file: Design-Implementation.tex
\section{Tool Support in Software Design and Implementation}
\label{sec:tool}
\subsection{Framework Support}
\label{sec:framework}
\begin{figure}
    \centering
    \includegraphics[width=0.6\linewidth]{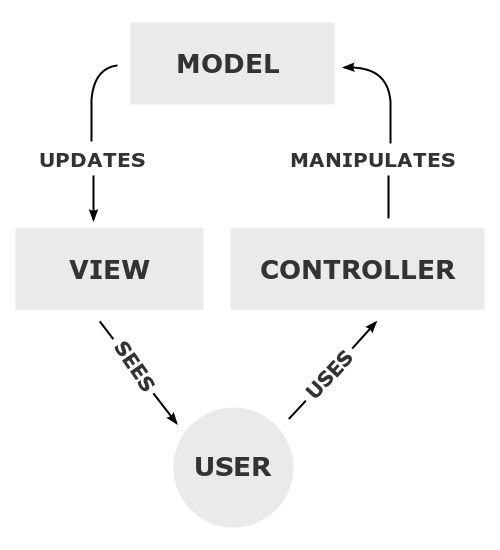}
    \caption{Architecture of MVC frameworks}
    \label{fig:mvc}
\end{figure}
Centralized access control mechanism are proposed as a way to tackle with access control vulnerabilities. As pointed out in several works~\cite{giffin2012hails, yang2016precise}, access control policies are usually strewn throughout the application code as guards, which are error-prone to be missed or wrongly implemented. Therefore, some previous works propose web frameworks that centralize access control and enforce the policies automatically based on the specification, like CanCan/CanCanCan~\cite{cancan} for Ruby applications and PHP-Casbin~\cite{caslin} for PHP applications.

Principle of least privilege is considered in splitting web applications into multiple components, thus enabling higher level of data access protection. Diesel~\cite{felt2011diesel} and OKWS~\cite{krohn2004building} provide web frameworks that follow this principle with explicit separation from the developers. Passe~\cite{blankstein2014automating} also tackle with unintended data leaks and unauthorized access by splitting applications into different components, but in an automatic way.  It applies three design principles in its design:
\begin{itemize}
    \item Split application code into components based on natural isolation boundaries.
    \item Apply the principle of least privilege.
    \item Automate the POLP split by dynamic analysis.
\end{itemize}
These principles can be applied in today's scale-out architectures more easily because developers now are encouraged to design and implement applications that provide narrowly defined interfaces. Therefore, Passe leverages this to automatically split applications and isolate different components. It runs the applications as a set of processes as 'views' (note that this is different from the 'view' in MVC architecture), and limit the data queries a view can make to restrict the information leak. Besides, Passe also captures the data and control-flow relationships between queries and other data in the system. This is learned from a trusted learning phase when the applications will run normally under the monitoring of Passe. Passe infers the relationships between database query results and following queries to detect the dependency between them like equality or set-membership. For example, in a social media system, a view displays all of a user's friends' could issue two queries on database:

\begin{lstlisting}[language=Python]
friends = fetchFriendsOf(uname = "Bob") 
updates = fetchUpdates(author in friends)
\end{lstlisting}

where the second query depends on the result of first query. This is a data dependency relationship. To catch this, Passe enforces this relationship when the second query only contain arguments from the first query. 

\begin{figure}
    \centering
    \includegraphics[width=\linewidth]{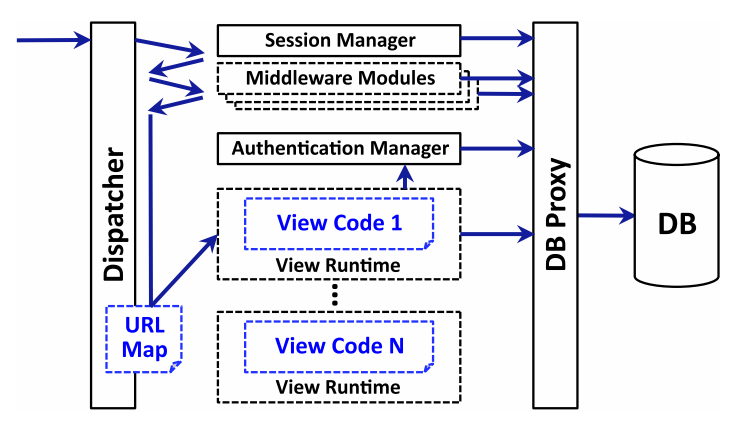}
    \caption{Architecture of Passe on MVC design patterns}
    \label{fig:passe}
\end{figure}
The author implemented Passe as a replacement of Django based on the MVC design patterns as shown in Figure~\ref{fig:passe}, which give a natural separation between computation and data. The blue solid arrows in the figure shows the communication channels in Passe's runtime. Two mechanisms in Passe are used to enforce the dependencies in communication channels: database proxy and cryptographic tokens. The tokens are encoded from results of previous queries and decoded by the proxy to make sure the data dependency is satisfied. Every time a query arrives, the proxy will lookup an access table and perform a two-stage check: First whether the requested query is allowed for the view that dispatch it. Second whether current constraints contained in the token satisfy the dispatched query.

\begin{figure}
    \centering
    \includegraphics[width=\linewidth]{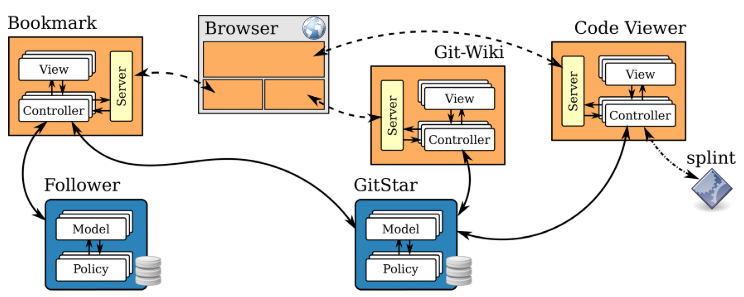}
    \caption{An example of Hails platform.}
    \label{fig:hails}
\end{figure}

While the above works try to find a compatible solution under the current framework, some other works address the internal drawback of MVC models and argue to revolutionise it. A widely adopted principle is the access control policy should be bound to application data so as to reduce the repetitive access checks in application code. CLAMP~\cite{parno2009clamp} enforces access control on user data with a trusted Query Restrictor. For each authenticated user, CLAMP verifies the identities then instantiates separate virtual web server instance for each user. W5~\cite{krohn2007world} proposes a novel model where users are in charge of their own data and specify policies alongside with their data. In their model, the logic code of applications is separated from data and security policies. Similarily, HAILS~\cite{giffin2012hails} extends the popular model-view-controller (MVC) models into a unique model-policy-view-controller (MPVC) models. In traditional MVC models, there is no inherent notion of security policy, so the access control policy is embedded into different layers, which is easy to be overlooked and introduce vulnerabilities. Hails makes the security policy as an explicit concern in their MPVC models. In their security model, trustworthy extensible web platforms run untrustful third-party apps that provided by other vendors. These third-party apps must manipulate sensitive user data from the platform to perform their functionalities. However, once the access permission to data are granted, it is hard to constrain what the app would do with this data holistically. MPVC models are proposed to target at this scenario. The components fall into two categories: 
\begin{itemize}
    \item \textbf{MP}: data model and policy logic. It interacts with databases and provides an API through with any other components access the data they want from databases under the policy enforcement. As shown in Figure~\ref{fig:hails}, Follower and GitStar are two MPs.
    \item \textbf{VC}: views and controllers. It is the part that interact with users and provide different functionalities, developed by the third parties. As shown in Figure~\ref{fig:hails}, those apps, Bookmark, Git-Wiki and Code Viewer, are VCs. Users directly interact with them through browsers.
\end{itemize}
To specify the security policy, developers write domain-specific language alongside the data model. In this security policy, the subjects are named as \textit{principals}, including four types: users that identified with user-names, remote web sites that identified with URLs, each VC, and each MP. 

\subsection{Database Support}
\label{sec:database}
Data protection policy is usually implemented in application layer as checks in application code. However, this approach could be error-prone because they are spread around the application code and easy to be missed. DBMS do provide some support for specifying fine-grained access control policy by SQL conditions~\cite{dewitt2004limiting} and views~\cite{rizvi2004extending, bender2013fine}. These solutions require specialized support within DBMSes. To make the database-level solution more portable, DataLawyer~\cite{upadhyaya2015automatic} and Qapla~\cite{mehta2017qapla} proposes to execute the query rewriting in middleware layer to enforce the access control policies. It allows the developers to express access control policies in SQL-like language. The difference between them is Qapla has an extra requirement on the developer-provided policies, which need to be indexed by columns to improve the efficiency. For example, in a fictitious company, the policy on names of employees can be expressed as:
\begin{lstlisting}[language=Python]
name :- EXISTS(SELECT 1 FROM Employees 
        WHERE empID = $user)
\end{lstlisting}
All the policy can be expressed as SQL WHERE clauses like this example. Qapla intercepts all queries and rewrite them to enforce the access control policy.

\subsection{Language Support}
\label{sec:language}
While access control checks act as gatekeepers when data are read or written, another way of protecting data against illegal access is to enforce the data flow to trusted sink. \textit{Information Flow Control} is the solution that can be used to achieve this. Many language-based techniques are explored to enforce the information flow. Jif~\cite{pullicino2014jif} extends Java with support for information flow control, enforced at both compiler time and runtime. Jeeves~\cite{yang2012language} is a Scala library that similarly enforces information flow policies. IFDB~\cite{schultz2013ifdb} works alongside decentralized information flow control languages, using tags attached to data to denote data sensitivity. Jacqueline~\cite{yang2016precise}, a Python web framework, uses multi-faceted execution to dynamically enforce security policies. LabelFlow~\cite{chinis2013practical} extends PHP bytecode interpreter to track security labels at runtime. LWeb~\cite{parker2019lweb} is another framework that marries the LIO Haskell IFC enforcement library to enforce label-based information flow policies.

The work we previously discussed in Section~\ref{sec:framework}, Hails also uses language-level confinement and enforces mandatory access control throughout the system. Security policies, referred as \textit{labels} in Hails, are written to specify which principals can read and write which piece of data, for example, $<alice \vee bob, alice>$. Data labeled with $<S,  I>$ can only be sent to a principal $p$ when $p\Rightarrow S$. The similar rules apply to writers. To maintain these policies, the trusted runtime will check whether they are still satisfied before permitting communication in any thread. In implementation of Hails, it spread the security enforcement across language-level, OS-level and browser-level. It crucially takes advantage of Haskell's type system, Linux isolation mechanisms and client-side sandbox. 

Storm~\cite{lehmann2021storm} is a web framework that targets at allowing developers to build applications in MVC architecture but with compile-time enforcement on security policies. The main idea is to take advantage of refinement types: a type endowed with a predicate that assumed to hold for any element of this type. In Storm, their unique insight is to specify the logical assertion as the security policies on certain piece of data. In this way, the access policies are enforced statically at compile-time and non-invasively. No labels or special APIs need to be invoked when developing. To apply this insight, Storm enhances the data model in MVC into a refined data model with annotations of security policies. To implement this, it takes advantage of LiquidHaskell~\cite{liquidhaskell}, which is an refinement type checker to verify whether the application code follows the enforcement of security policies. In Storm, the authors proves that the Storm API is secure and cannot leak or corrupt sensitive data.

To implement Storm, two foundational blocks are involved: refinement types and compositional IFC. Refinement types decorate a existing type with logic assertions to specify a subset of values of the original type. By writing pre- and post-conditions, the user can refine the input and output of functions. With bounded refinements, developers can abstract the policies and invariants of applications. An SMT solver can be used to automatically verify all the refinement variables can be instantiated. Compositional IFC is forced in Storm as follows: Every applications are regarded as a set of handlers. Every handlers are regarded as a sequence of operations. (e.g. $e_12 = do e_1;e_2$, where $;$ is sequential composition operator). Each data operations are regarded as read sensitive data with authorizees or write data to observers. Then the information flow control is enfoced as if operation $e_i$ reads data as authorizees $auth_i$, any subsequent operation $e_j$ can only write to observer $obs_k\subseteq auth_i$. To prevent path explosion, a two-step compositional approach is applied here. First, every primitive operations are assigned with types that contain the corresponding authorizees and observers. Next, the operator $;$ is assigned with a type that enforces that for any two composed computations $e::<auth, obs>$ and $e'::<auth', obs'>$, for the output $e;e'::<auth'', obs''>$we have:

\begin{gather*}
    obs' \subseteq auth\\
    auth''=auth\cap auth'\\
    obs''=obs \cup obs'
\end{gather*}
In this way, the IFC constraints are enforced.

Storm is the first framework that statically enforce security policies in language level. The advantage of this solution is: First, it does not impose any additional run-time overhead as other runtime checking mechanisms. Second, the potential violations to security policies are detected in the early stage before deployment, which align with our consideration in Section~\ref{sec:background} that access control vulnerabilities are better to be caught before deployment. Third, no additional developers' effort on the underlying enforcement mechanism is explicitly required in this context. 

The limitation of language support is that it only applies to language implementations with certain features, while most web applications are written and run in mainstream language supports. Besides, these solutions are not free of any additional effort. In fact, in the experiment of Hails, developers felt the difficulty on implementing MPs. Though a DSL could make the process easier and help developers to understand the underlying enforcement, it still requires additional effort compared to normal development process. 
In Storm, The programmers still need to write a refined models file, which is a centralized place to specify the security policies.

Aside from information control flow, RESIN~\cite{yip2009improving} is a runtime prototyped in PHP that allows the developers to specify data flow assertions and associate them to application data. It checks the data assertions when it flows cross boundaries such as writing data to files or network. 

\subsection{Development Environment Support}
\label{sec:development}
Zhu et al.~\cite{zhu2015mitigating} propose a hybrid approach to detect access control vulnerabilities based on interactive static analysis. They firstly conduct a comparative study on six PHP web applications, which focus on Security Sensitve Operations  (SSO) that are database operations. They manually examined and identified access control checks in source code of these applications, then summarize the distribution of access control patterns in these applications. From this, they observe that access control patterns usually have low repetitions, which means it is relatively hard to construct pattern models and detect them automatically. Also, they find there is much noise in code instances of access control code, which further affect the accuracy of auto-detection methods, and results in false positives in these solutions. 

\begin{figure}
    \centering
    \includegraphics[width=\linewidth]{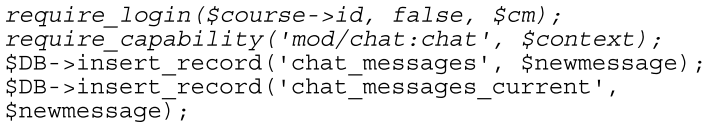}
    \caption{Access control on Moodle database SSOs}
    \label{fig:aside-1}
\end{figure}

\begin{figure}
    \centering
    \includegraphics[width=\linewidth]{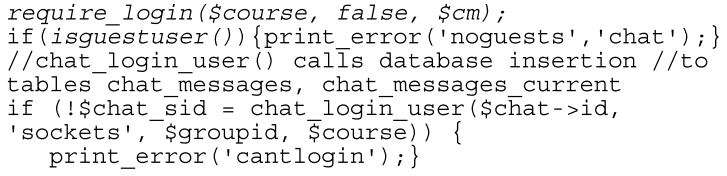}
    \caption{Access control vulnerabilities detected in Moodle by ASIDE}
    \label{fig:aside-2}
\end{figure}

Therefore, in their work, they seek for developer's input to identify access control models with a relatively little distraction to developers' normal workflow. Their interactive tool ASIDE identifies SSO as developers write code in IDEs, then place a yellow notification alongside as a remind, which ask developers to highlight access control logic so the warnings can be removed. Meanwhile, ASIDE uses static analysis to help find the incorrect access control logic. For example, as shown in Figure~\ref{fig:aside-1}, ASIDE identifies SSOs of INSERT to \textit{chat\_messages} and \textit{chat\_messages\_current} and the surrounding access control logic. But when in Figure~\ref{fig:aside-2}, similar SSOs are performed in function \textit{chat\_login\_user} but without similar protection as the previous code. Therefore, ASIDE alerts the developers in case there is a potential vulnerability. 

%% file: Analysis-Testing.tex
\section{Software Analysis and Testing}
\label{sec:analysis}
\subsection{Outlier Detection}
\label{sec:outlier}
As a type of semantic bugs, detecting access control vulnerabilities require a ground truth of the correct semantics. Then the tools detect vulnerabilities candidates based on the outliers to the ground truth.  To obtain the ground truth, the most direct way is to require the developers to express their intended access control policies. However, this can be a burden to developers. Therefore, researchers came up with workaround solutions to mitigate the additional effort. 

MACE~\cite{monshizadeh2014mace} only require annotations on the variables that correspond to user ids, roles and other session-related variables. It depends on program analysis and symbolic evaluation to identify the \textit{authorization context} for web applications from these annotations. It then checks the authorization context with every program point if they access the same resource from different execution paths. If these authorization contexts are mismatched, there is possibly a potential access control vulnerabilities. MACE infers and expresses access control rules in a four tuple <U, R, S, P>, where U is the authenticated users, R is the set of roles, S is the session variables or identifier, and P is the permissions required on the resources. It then computes the authorization context by control flow and data flow analysis for every database queries in the application code. It firstly builds a source-sink graph based on data dependency of the programs, then collect the constraints in sinks to build up the authorization context of each query. In this way, MACE could detect both vertical privilege escalation and horizontal privilege escalation. Vertical privilege escalation happens when users in low privilege level could access to resource of high-privilege users. Horizontal privilege escalation happens when users can access the resource of another user that is from same privilege level.

\begin{figure}
    \centering
    \includegraphics[width=\linewidth]{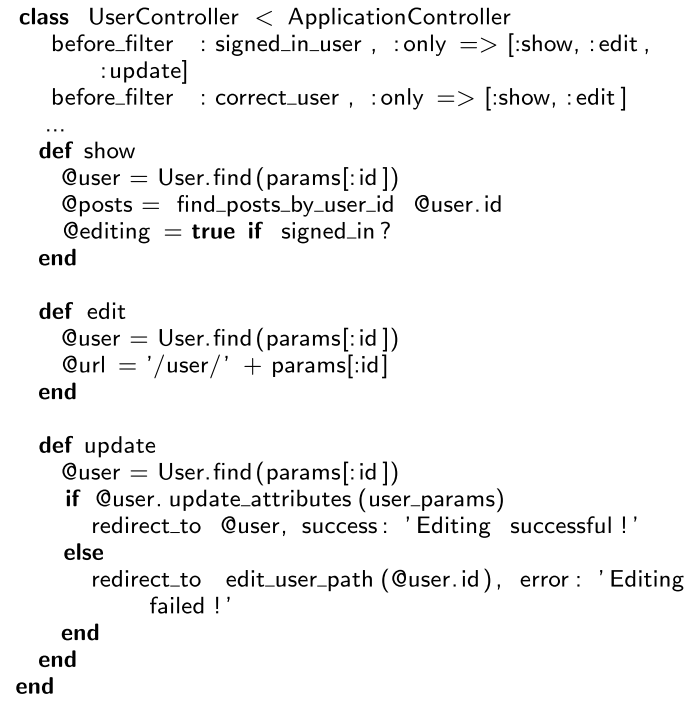}
    \caption{An access control bug from MediumClone}
    \label{fig:example-space}
\end{figure}
SPACE (Security Pattern Checker)~\cite{near2016finding} also claims to be a specification-free tool for finding access control bugs. How it avoid specifications from developers is to make the observation that developers usually applied similar patterns of checks for a given resource type. Therefore, developers only need to provide mapping to their selected catalog of access control patterns from SPACE. Then SPACE checks whether each data exposure is allowed. SPACE comes up with six patterns as `catalog':
\begin{itemize}
    \item Ownership. 
    \item Public Objects.
    \item Authentication.
    \item Explicit Permission.
    \item User Profiles.
    \item Administrators.
    \item Explicit Roles.
\end{itemize}
SPACE uses symbolic execution to extract the data exposures from source code. Then it uses a constraint solver to check whether the constraints associated with the data exposures are also in the allowed scope of user-given catalogs.

We illustrate the workflow of SPACE with a example from MediumClone. Figure~\ref{fig:example-space} shows an access control bug from this application. In the \textsc{UserController} class, the \textsc{edit} action will show a page of editing the profile and click on buttons will result in submitting the updated profile form, which results in the invocation of \textsc{update}. However, the filter of access control only apply to \textsc{show} and \textsc{edit}, which means an attacker can directly bypass the edit pages and send the post request to \textsc{update} handler, then update other users' profiles. This is regarded as an access control bug in MediumClone. To detect this bug, in SPACE, the user will specify the access control policy in form of role-based access contrl (RBAC) models as:
\begin{lstlisting}[language=Python]
Space.analyze do
    mapping User: RBACUser, 
        Post: OwnedObject(user: owns)
end
\end{lstlisting}

which means only the owner of a user profile can update the information. SPACE then builds the exposure relation of \textsc{update} as \textsc{(update(user.profile), true, write, user)}. By substituting the RBAC rules SPACE can determine whether illegal access is allowed in this exposure, thus detects an access control vulnerability. 

\begin{figure}
    \centering
    \includegraphics[width=0.8\linewidth]{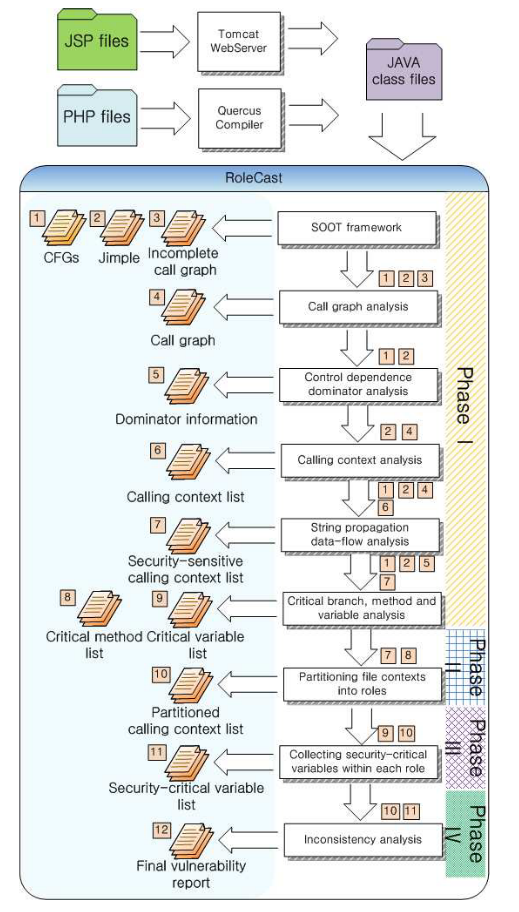}
    \caption{Architecture of RoleCast}
    \label{fig:rolecast}
\end{figure}

Some other works push hard to be specification-free. Instead, they construct their ground truth from the code implementations and detect the outliers that are deviated from their peers. RoleCast~\cite{son2011rolecast} is such a static anlaysis tool that does not rely on any annotations from developers or security policy specifications. Instead, it tries to catch some common patterns of security checks in web applications and apply to infer the missing checks. The architecture of RoleCast is shown in Figure~\ref{fig:rolecast}. It works in four phases. First, it analyzes interprocedural control flow to gather the context of sensitive events, including the critical variables in conditions. To eliminate impossible candidates, it filters out those without abnormal exit branches because those conditions are possibly not related to security checks. Second, RoleCast divides program files into different groups based on the roles they are attached to. It partitions them by analyzing the file contexts and minimize the number of files that shared between different roles. Third, RoleCast will determine for each role what is the set of critical variables to check. Last, RoleCast will check the calling contexts of each sensitive event and reports potential vulnerable cases if they match one of the following patterns: (1) calling contexts of a security-sensitive operations do not have any check; (2) the role contains a unique context that is not able to check the consistency; (3) A check is inconsistent with the majority of other calling contexts.

Another work by Sun et al.~\cite{sun2011static} also detect access control vulnerabilities without requiring any explicit access control specifications from developers. They catch the \textit{implicit access control assumptions} from PHP programs source code where it implicitly indicates the intended accesses of every roles in the system. Therefore, if a role can force browsing a privileged web page in this system, it is potentially a vulnerability. In their work, they abstract the web application systems with $k$ nodes as follows:
\begin{equation*}
    P_r=(S_r, Q_r, E_r, I_r, \Pi_r, N_r)    
\end{equation*}
where $S_r$ is the entry set of this application for role $r$. $Q_r$ is the set of sensitive application states that a web page $n_i$ contains for role $r$. $E_r$ is the set of explicit edges between pages. $I_r = {<ni, nj> | 1 \leq i,j \le k}$, where $n_j$ is the set of pages that can be forced browsing to from page $n_i$ under state $q_i$. $\Pi_r$ is the set of all web pages that role $r$ can navigate to. $N_r$ is the set of web pages that are explicitly reachable through $E_r$, starting from $S_r$. In this way, let $a,b \in R$ are two roles in this application, this work defines access control vulnerability as:
\begin{equation*}
    n \in N_a \land n \notin N_b \land \exists \pi_b \in \Pi_b (n \in \pi_b)
\end{equation*}
This definition means, if a page legally accessed by role $a$ can be forced browsing by role $b$, but it cannot be explicitly accessed by $b$, this page is potentially an access control vulnerability. 
\begin{figure}
    \centering
    \includegraphics[width=\linewidth]{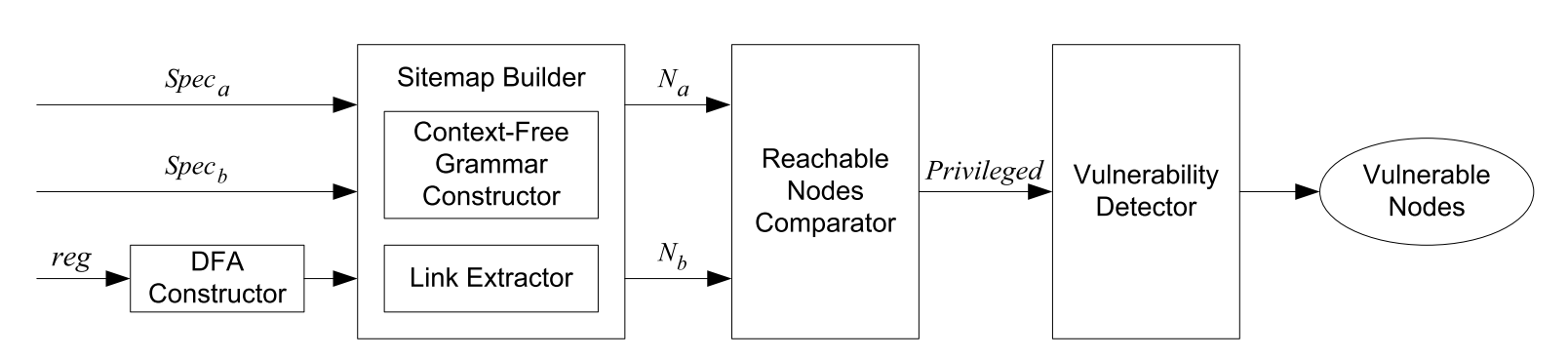}
    \caption{Architecture of Sun et al.'s Work}
    \label{fig:sunetal}
\end{figure}
The workflow of this work is shown in Figure~\ref{fig:sunetal}. The only input required from developers is the entry set $S_r$ and the state set $Q_r$. The tool builds sitemaps from the entry set by control flow analysis and find the branches that a role can trigger. Then it collects recursively all the PHP files that can be triggered by each roles. To resolve the infeasible path, this tool uses Z3 to solve constraints. With the link extractor, this tool can construct per-role sitemaps. Next, with collected $N_a$ and $N_b$ set, the comparator will compare between two sets and infer privileged nodes, which mean only privileged users can access. It then uses analyzer to emulate forced browsing attempts on these privileged nodes from low-privilege users to determine whether they will be successful. Finally, with manual confirmation, this tool will output the vulnerable nodes.

Aside from static methods, dynamic methods can be applied to construct the access control models of web applications, for which their assumption is normally the accesses in the applications follow the access control policy and can serve reversely as specifications. Waler~\cite{felmetsger2010toward} use dynamic analysis to infer a set of behavioral specifications, then use model checking to identify paths in programs that are likely to break the specifications as vulnerabilities. CanCheck~\cite{bocic2016finding} takes advantage of the applications of the access control library CanCan~\cite{cancan} to help instrument authorization checks in web applications, thus being able to construct access control models from the dynamic execution, since instrumented execution is made easier with the centralized CanCan modules and explicit \textit{can?} checks. Then it infers formal access control models into a set of first order logic formulas. If existing an action, for certain operations on a certain object, the authorization checks are not implemented as specified by the access control models, CanCheck regards it as an access control bug. 
\begin{figure}
    \centering
    \includegraphics[width=\linewidth]{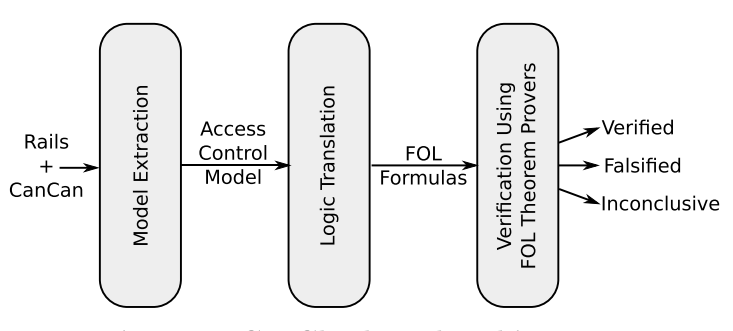}
    \caption{Workflow of CanCheck}
    \label{fig:cancheck}
\end{figure}

FixMeUp~\cite{son2013fix} goes a further step to not only detect access control bugs, but also generate repairs for them. The unique observation it makes is that the missing access control checks are always presented in the call context so that it can reuse the statements to generate repairs. The process of how FixMeUp works is as follows. First, it requires input of high-level specification on access control policy, which could obtain either by user manually written or from program analysis. Typically, a developer can annotate in the source code to mark the correct access control checks, the sensitive operations, and the user role that this policy applies. Then, FixMeUp generates access control templates(ACT) based on the specification. It starts with the correct access control checks and compute in backward interprocedural manner to build the representation of concrete low-level policy specifications. And this representation will also serve as program transformation template when later generating fixes.
Next, FixMeUp checks every calling context to verify whether the access control logic matches the ACT. To identify the relevant statements, it looks at the entry points of programs, where the access control checks are usually stylized. It then generates the repairs if there is missing checks.

\subsection{Automated Testing}
\label{sec:testing}
Automated testing for access control in web applications is a more difficult task compared to analysis approaches. The difficulty comes from the generation of test cases, which includes a series of user interactions, and test oracle, which largely depends on business logic of the applications. For most web applications where the access control components are not unified in special format like XACML, and the source code is not available, the access control testing have to be end-to-end. The tools send out web requests and analyze the responses to decide whether the test is passed or failed. 

AuthScope~\cite{zuo2017authscope} tackles with the test case generation by using differential traffic analysis to compare and recognize the fields of interest in web requests, then substitute the field in requests and observe the server response. This work performs a large scale study based on their detection. To scale their method, the first challenge is how to pass the authentication of most apps so that the tool can obtain post-authentication requests and responses. This is solved by using social login interface of these apps (for example, Facebook login). Second, they need to identify the field that are possibly guessed to generate the test cases. To solve this, they use Euclidean distance measure the randomness of identifiers. Third, how to confirm a vulnerability is exploited from the responses (i.e. test oracle generation). They similarly apply differential analysis to response messages between different accounts and filter out trivial differences like time stamps in the responses.

LogicScope~\cite{li2013logicscope} catches logic vulnerabilities within web applications, among which unauthorized access is a major category. They model the logic of a web application using finite state machine (FSM) over the observed user inputs from execution traces, in which the users follow the legal navigation paths. Then based on the FSM they construct, they make up unexpected inputs for each state and evaluate the corresponding responses to see whether this input reveals a logic vulnerability.

\subsection{Formal Method}
\label{sec:formal}
Ur/Web is a domain-specific language for web application development. It contains a static analysis tool UrFlow~\cite{chlipala2010static}, which asks for security policy specifications in the form of SQL queries, then use symbolic evaluation and automated theorem-proving to statically check these policies. 

Wang et al's work~\cite{wang2013explicating} considers the authorization problems in SDKs. In modern applications, developers could take advantage of online providers to do authentication and authorization services. However, how to use these SDKs in a secure way require the developers to be aware of assumptions in both SDKs and the underlying runtime systems, which is difficult for most of the developers to understand. In their work, they try to demonstrate an explication process in which they devise precise definitions of desired security properties to construct the security models of the SDK and the underlying systems. Then they apply formal verification to detect counterexamples so that they can either find the incorrect part of their model, or add new assumptions to their models. In this iterative process, they are able to get a set of assumptions and models, which can be enforced when the SDK is used by other third-party developers. They advocate this explication process as part of the engineering process of SDKs.

\begin{figure}
    \centering
    \includegraphics[width=\linewidth]{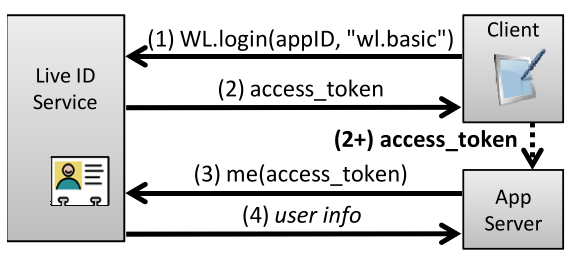}
    \caption{An example of misusing SDKs. The access token from Live ID service is wrongly sent to a third-party server, which could be exploited to stole sensitive information by malicious apps.}
    \label{fig:sdk}
\end{figure} 
In their explicated assumptions, they define their security properties as follows: the granularity of their security model is session. The basis of security model is the usage of secrets and signed data, which are used to idenfity each users. The desired security goals include authentication violations, authorization violations and association violations. All the assumptions are built based on these properties.

%% file: Runtime-Monitoring.tex
\section{Runtime Monitoring}
\label{sec:monitor}
In Section~\ref{sec:language}, we discuss the language supports of information control flow, which can be enforced during runtime. Aside from these works, there are other works trying to enforce access control during runtime without special requirements to the existing code bases. Nemesis~\cite{dalton2009nemesis} integrates authentication information and access control policy specified by developers to make sure only certain authenticated users are able to access the legal resource. To trace how the authentication information flows, Nemesis takes advantage of the insight that most web applications use similar design of storing usernames and passwords. Therefore, Nemesis applys Dynamic Information Flow Tracking to track user credentials. It assigns an additional shadow HTTP cookie to a authenticated user so it can follow the subsequent requests sent by the same user. 

\begin{figure}
    \centering
    \includegraphics[width=\linewidth]{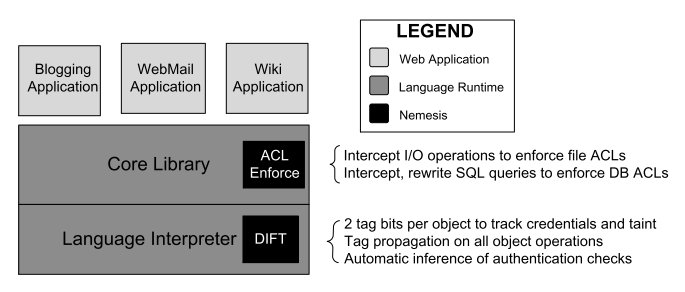}
    \caption{Architecture of Nemesis}
    \label{fig:nemesis}
\end{figure}

To track the authentication information, Nemesis uses DIFT and keeps two taint bit for each data items, a credential bit and a user bit. Besides, Nemesis requires developers to tag credential data retrieved from database. When a data item is tagged with a taint bit and compared to credential data, Nemesis regards it as an authentication operation. If the two operands are the same, it records this client as authenticated. To deal with user registration, Nemesis infers by a SQL INSERT statement. 

FlowWatcher~\cite{muthukumaran2015flowwatcher} detects access control vulnerabilities during runtime. It makes the observation that most web applications use similar simple access control models, which can be described in a rule-based specification as user-data-access policy, namely UDA policy. This policy specification allows dynamic tracking and monitoring of access control vulnerabilities once violations of the UDA policy happens. In this way, an enforced policy can be created and reused across different applications, which contains less code and is easy to be implemented correctly. Besides, it requires no modifications to the original applications.

To apply this approach, there are two challenges. First, how to make it easy for developers to express the access control policy and keep it updated when new roles and objects are added to the application. This work applies UDA policies to solve this problem. The UDA policy are specified through two entities: data objects, and users and groups. And the rules in UDA policies can be two types: 1) Definition/removal rules, which update access control policy when there is objects creation or deletion. 2) Update rules, which related to updating on the existing object. It matches an HTTP request to corresponding users and resource based on its URL, header and body. Second, how to efficiently track user data with little performance overhead. FlowWatcher uses shadow policy state to avoid the delay of service performance. It checks the request URL, bodies and the response of this request to see whether it requires updating of UDA policy. FlowWatcher tracks unique data in the system by matching the data value and monitors the value in HTTP response. 

%% file: discussion.tex
\section{Open Problems}
\label{sec:discuss}
Access control vulnerabilities are not newly emerged. As long as there is data and data access inside web applications, there could be access control vulnerabilities. However, with the growing dependency of society on web services and the blossom of IT industry, this problem is becoming even prevalent and severe. Following trends have emerged:
\paragraph{Heterogeneous Interfaces in Web Applications} With the development and evolution of web services, nowadays web applications are providing heterogeneous APIs, including GraphQL API, RESTful API, xmlRPC APi, etc. This also aligns with the demand of cross platform applications. These APIs share similar functionality and same user data with the main applications. Any exploitation of these APIs could result in data leak as severe as the main web applications.
\paragraph{Data Hosted on Third-party Platform} Cloud hosting provides a scalable and friendly management solution for web application developers. However, this also introduces new risk of data leak. Data leak from third-party platform is not only a configuration problem as already studied by previous work~\cite{zuo2019does}. Interactions to third-party platform also increase the complexity of code implementation. In a real world example from Bug Bounty programs~\cite{fetlife}, a Google API Key is leaked in front-end code of this web applications, resulting in access control vulnerabilities.
\paragraph{Emerging Software Development Style} Tech companies are pushing hard to make the software development more approachable. Even novice programmers can set up and tailor the web application based on website builders like WordPress, Joomla, Drupal, etc. A exposed access control vulnerabilities in such software will result in many websites under the risk due to software reuse. Another new trend is serverless computing, where the data flow inside a web applications could even not be under total control of developers, making current approaches like information flow control not applicable in these scenarios.
\paragraph{Adoption of Machine Learning Models} Applying machine learning models in web applications means the user data not only flows through application code, but also potentially flows and leaked from the model parameters, which could also become candidates of illegal access attacks and harm users' privacy.

Aside from these new trends that increase the complexity of access control vulnerability problem, even for the most simple and original form of this problem, it is still far from a perfect solution. Based on the importance and severity of access control vulnerabilities, we advocate to solve it in early stage of software development. Since in essence they are semantic bugs, we believe that to prevent and detect access control vulnerabilities as early and thoroughly, the participation of developers is crucial. And the key problem here is how to make it developer-friendly to express and perceive potential risk: 

\paragraph{Access Control Configurationize} 
Access control is more easily to be vulnerable when it spreads across the code base. As many previous works propose, a centralized access control module would definitely reduce the chance of access control vulnerabilities and increase the auditability of access control implementations and policies in a system. With this centralizing design, access control will move towards more of a configuration problem where the developers will focus on what access control configurations should be placed on the interfaces of a system. And how they express the access control enforcements will also depend more on the specific configuration language instead of the language they use to implement the system.

\paragraph{Improving the Access Control UIs}
Another direction to help developers avoid access control vulnerabilities is to help them know clearly about their implementation, so that they can easily find the holes in their current design. The de facto access control is still expressed by application code or text files. However, this kind of expression is abstract and cannot help developers a lot when the access control policies become complex. Therefore, building an informative and handy access control UIs could be non-trivial to help developers. Even simply expressing access control in a visualized interface is a not a simple problem. First, there could be a lot of subjects and objects in the system. Plainly displaying them and building up the access control matrix cannot help developers but will in reverse impede them from identifying the actual problem. Second, access control UIs need to reflect the change when developers introduce new endpoints and new authorization checks. How to extract them without much additional effort from developers also presents a problem.

\paragraph{White-box Access Control Testing}
In the existing literature, researchers explore a bunch of approaches in black-box access control testing to identify the vulnerabilities. However, black-box testing is insufficient in solving this problem. In a typical web applications, there are orphaned pages and operations that are not exposed to outside publicly. Black-box testing need to guess and hit them by chance, which will take much computation resource. With source code and more information, the hidden part can get more thorough testing without the need to guess through the random space. And the penetration testing can be much more efficient. Currently, the white-box testing is carried out by security experts and requires manual assessment on the web application code. This brings additional financial burden to companies and cannot guarantee the coverage in a systematic way. Therefore, we encourage the community to summarize experience of white-box penetration testing and automatic this process.

%% file: conclusion.tex
\section{Conclusions}
In this article, we discuss the characteristics of access control vulnerabilities in web applications, the challenges of eliminating access control vulnerabilities from web applications, and the state-of-the-art approaches in different stages of software development to deal with these vulnerabilities. Based on our study on the literature, we discuss the open problems in this field. We hope this report can encourage further research toward the elimination of access control vulnerabilities.